\documentclass[sigconf]{acmart}

\usepackage{soul} 
\sethlcolor{yellow} 
\usepackage{xcolor} 
\usepackage{booktabs} 
\usepackage{array} 
\usepackage{colortbl} 
\usepackage{enumitem} 

\AtBeginDocument{%
  \providecommand\BibTeX{{%
    \normalfont B\kern-0.5em{\scshape i\kern-0.25em b}\kern-0.8em\TeX}}}

\copyrightyear{2024}
\acmYear{2024}
\setcopyright{rightsretained}
\acmConference[UIST '24]{The 37th Annual ACM Symposium on User Interface Software and Technology}{October 13--16, 2024}{Pittsburgh, PA, USA}
\acmBooktitle{The 37th Annual ACM Symposium on User Interface Software and Technology (UIST '24), October 13--16, 2024, Pittsburgh, PA, USA}
\acmDOI{10.1145/3654777.3676449}
\acmISBN{979-8-4007-0628-8/24/10}

\newcommand{\rev}[1]{\textcolor{black}{#1}}

\begin{document}

\title{CookAR: Affordance Augmentations in Wearable AR to Support Kitchen Tool Interactions for People with Low Vision}

\author{Jaewook Lee}
\affiliation{%
  \institution{University of Washington}
  \city{Seattle}
  \state{WA}
  \country{USA}
}

\author{Andrew D. Tjahjadi}
\affiliation{%
  \institution{University of Washington}
  \city{Seattle}
  \state{WA}
  \country{USA}
}

\author{Jiho Kim}
\affiliation{%
  \institution{University of Washington}
  \city{Seattle}
  \state{WA}
  \country{USA}
}

\author{Junpu Yu}
\affiliation{%
  \institution{University of Washington}
  \city{Seattle}
  \state{WA}
  \country{USA}
}

\author{Minji Park}
\affiliation{%
  \institution{Sungkyunkwan University}
  \city{Suwon}
  \country{Korea}
}

\author{Jiawen Zhang}
\affiliation{%
  \institution{University of Washington}
  \city{Seattle}
  \state{WA}
  \country{USA}
}




\author{Jon E. Froehlich}
\affiliation{%
  \institution{University of Washington}
  \city{Seattle}
  \state{WA}
  \country{USA}
}

\author{Yapeng Tian}
\affiliation{%
  \institution{University of Texas at Dallas}
  \city{Richardson}
  \state{TX}
  \country{USA}
}

\author{Yuhang Zhao}
\affiliation{%
  \institution{University of Wisconsin-Madison}
  \city{Madison}
  \state{WI}
  \country{USA}
}


\renewcommand{\shortauthors}{Lee et al.}
\begin{abstract}
Cooking is a central activity of daily living, supporting independence as well as mental and physical health. However, prior work has highlighted key barriers for people with low vision (LV) to cook, particularly around safely interacting with tools, such as sharp knives or hot pans. Drawing on recent advancements in computer vision (CV), we present \textit{CookAR}, a head-mounted AR system with real-time object affordance augmentations to support safe and efficient interactions with kitchen tools. To design and implement CookAR, we collected and annotated the first egocentric dataset of kitchen tool affordances, fine-tuned an affordance segmentation model, and developed an AR system with a stereo camera to generate visual augmentations. To validate CookAR, we conducted a technical evaluation of our fine-tuned model as well as a qualitative lab study with 10 LV participants for suitable augmentation design. Our technical evaluation demonstrates that our model outperforms the baseline on our tool affordance dataset, while our user study indicates a preference for affordance augmentations over the traditional whole object augmentations.
\end{abstract}



\begin{teaserfigure}
  \includegraphics[width=\textwidth]{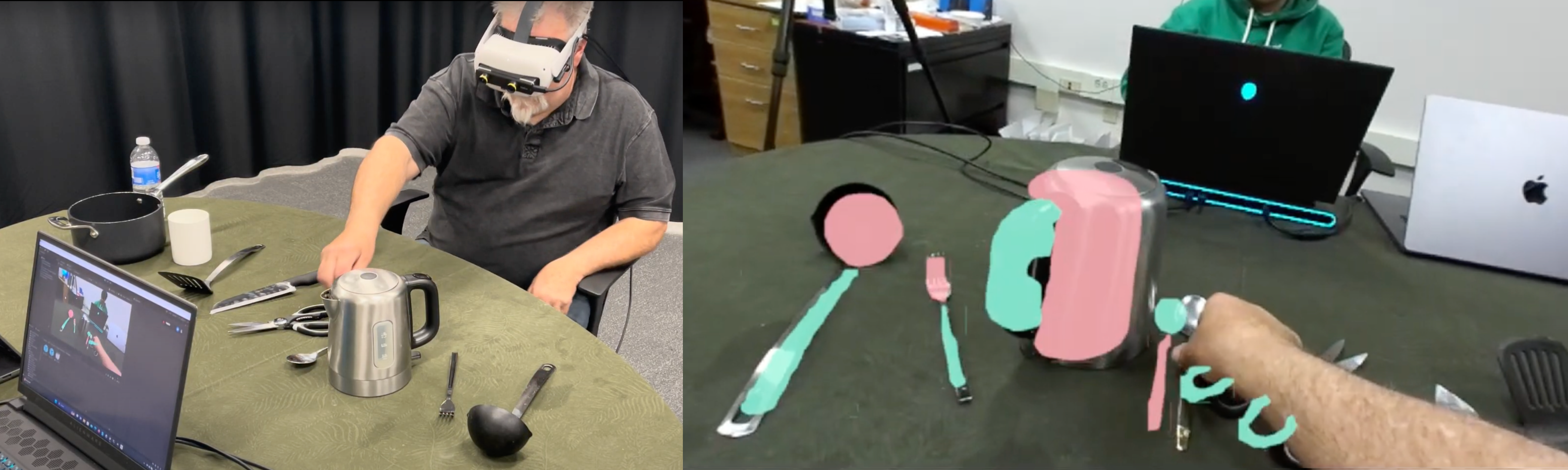}
  \caption{CookAR provides real-time object affordance augmentations in head-mounted AR to support cooking interactions. (A) a low vision participant uses CookAR to locate and grab a spoon; (B) the view in CookAR where kitchen tool affordances (grabbable \textit{vs.} hazardous areas) are recognized and augmented by solid-colored overlays, with green overlays for grabbable areas such as handles and red for hazardous areas such as a knife blade or the hot part of a tea kettle.}
  \label{fig:teaser}
\end{teaserfigure}


\maketitle

\section{Introduction}
Cooking is an essential activity of daily living, supporting independence~\cite{Miyawaki2012, Mynatt2001, taylor2016does} as well as both mental and physical health~\cite{Mills2017, Bilyk2009, Jones2019, taylor2016does}. However, cooking also involves significant visual tasks that can be challenging or dangerous for blind and low vision (BLV) people, especially when interacting with kitchen tools, such as sharp knives or hot pans~\cite{Bilyk2009, Jones2019, Li2021, Li2024, Wang2023}.

Unlike those who are completely blind, people with low vision (LV)---vision loss that cannot be corrected using glasses or contact lenses~\cite{Corn2010}---often rely on their residual vision in daily activities and use different low vision tools to enhance visual information~\cite{Szpiro2016a, Szpiro2016b}. With recent advancements in AI-powered augmented reality (AR), researchers have explored new possibilities for supporting LV individuals by automatically recognizing their environment and providing appropriate visual augmentations, including stair navigation~\cite{Zhao2019}, visual search~\cite{Zhao2016}, and sports~\cite{Lee2023}. While promising, these previous AR systems focus primarily on understanding effective augmentation designs~\cite{Zhao2019}, often oversimplifying the computer vision (CV) recognition in their development, thus neglecting the effects of technological limitations (\textit{e.g.,} CV inaccuracies and system delays) on user experience. Moreover, though there has been substantial formative work in BLV cooking within the HCI literature~\cite{Li2021, Li2024, Wang2023}, no previous AR system has been built to address LV cooking specifically.

We introduce \textit{CookAR}, a wearable stereo AR prototype that recognizes and augments cooking tool \textit{affordances} in real-time to support LV meal preparation. In contrast to prior AR research that highlights objects as a whole~\cite{Zhao2016,Fox2023}, we distinguish and augment the \textit{object affordance} specifically (\textit{i.e.,} component parts that afford interactions), such as the safe-to-handle ``\textit{grabbable}'' areas and the dangerous-to-touch ``\textit{hazardous}'' areas (Figure~\ref{fig:teaser}). To enable accurate affordance recognition, we constructed a custom egocentric image dataset for kitchen tool affordances by selecting and labeling images from the \textit{Epic Kitchens} dataset~\cite{Damen2018} and fine-tuned an affordance segmentation model. We then combined the \textit{ZED Mini}\footnote{https://store.stereolabs.com/products/zed-mini} stereo camera and an \textit{Oculus Quest 2}\footnote{https://www.meta.com/quest/products/quest-2/} headset \rev{to achieve a video passthrough AR system with CV and stereo depth estimation capabilities to precisely overlay affordance augmentations on the 3D environment in near real-time.}

To evaluate CookAR, we conducted a technical evaluation of our fine-tuned model as well as a three-part qualitative lab study with 10 LV participants. For the model assessment, we found that our fine-tuned affordance segmentation model (mAP of 46.3\%) outperformed the base RTMDet~\cite{Lyu2022} model (mAP of 12.3\%) in tool affordance recognition and segmentation. For the three-part user study, LV participants were first asked to locate and pick up cooking tools \rev{across three conditions: (1) with their typical method in daily life such as wearing corrective lenses (\textit{i.e.,} real-world baseline); (2) with CookAR displaying whole object augmentations (\textit{i.e.,} augmentation baseline); and (3) with CookAR displaying affordance augmentations.} They then completed a free-form cooking task with CookAR (Part 2) and brainstormed desired augmentation designs using design probes (Part 3). Findings indicate that participants prefer affordance augmentations over whole object augmentations in a kitchen as they enable faster understanding of an object's spatial arrangement and safe interaction parts. Most participants preferred affordance augmentations consisting of green solid overlay on grabbable areas and red outlines on hazardous areas. Moreover, participants identified five additional tool affordances with desired augmentations, including \textit{entry} (\textit{e.g.,} cup rim), \textit{exit} (\textit{e.g.,} carafe spout), \textit{containment} (\textit{e.g.,} cup base), \textit{intersection} (\textit{e.g.,} knife blade on butter), and \textit{activation} (\textit{e.g.,} carafe buttons) areas---all which should be outlined in a contrasting color (\textit{e.g.,} black or white).

In summary, our research contributions include: (1) CookAR, a fully-functional AI-powered wearable AR prototype that augments kitchen tool affordances for low vision users to enable safe and efficient tool interactions; (2) an egocentric affordance dataset for kitchen tools and an accompanying fine-tuned affordance segmentation model. \rev{To enable others to build off our work, this dataset and model are open-sourced at: \textcolor{blue}{\url{https://github.com/makeabilitylab/CookAR}}}; and (3) user study results with 10 LV participants that reveal user experiences with CookAR, preferences for augmentation designs, and five newly desired affordance areas.

\section{Related Work}
Our work builds on prior formative studies on low-vision cooking, wearable AR for accessibility, and affordance segmentation.

\subsection{Challenges in Low Vision Cooking}
People with low vision (LV) face challenges in everyday activities, such as cooking~\cite{Bilyk2009, wang2023characterizing, Wang2023, Li2021, Li2024}, shopping~\cite{Szpiro2016b}, navigation~\cite{zhao2018looks,Szpiro2016b}, and sports~\cite{Lee2023, rector2015exploring}. Among these, cooking is an essential task for an independent and healthy life~\cite{taylor2016does}. However, this task also poses major accessibility barriers and safety concerns to BLV individuals, including interacting with sharp knives and hot pans~\cite{kashyap2020behaviors,kim2022vision}. Consequently, they tend to eat more pre-processed food or frequently dine at restaurants, which can negatively impact their health~\cite{Jones2019, montero2005nutritional}.

To better understand how BLV people engage in cooking tasks, prior work has conducted both interview and observational studies \cite{Li2021,Lin2014,wang2023characterizing,Wang2023}. For example, Jones \textit{et al.}~\cite{Jones2019} surveyed 101 BLV participants in the U.K. about their shopping and cooking experiences, revealing that vision loss made cooking difficult and that the level of difficulty was correlated to the severity of visual impairments. Li \textit{et al.}~\cite{Li2021} analyzed 122 YouTube videos of BLV people preparing meals and interviewed 12 BLV participants about their cooking experiences. They identified several cooking-related challenges, such as utilizing cooking tools and tracking object dynamics in the kitchen. A follow-up contextual inquiry study~\cite{Li2024} examined how BLV people recognize cookware and utensils and measure ingredients. This study highlighted essential cooking-related information to convey, such as position, safety, and orientation of objects. 

Specifically for LV people, Wang \textit{et al.}~\cite{wang2023characterizing} conducted a contextual inquiry study, observing and comparing the cooking experiences between six LV participants and four blind participants. They found that while blind participants relied on touch, LV people used their vision extensively while cooking. However, compared to blind people, LV people felt less confident, less safe, and more tired and stressed due to their reliance on impaired vision. Moreover, LV individuals were less satisfied with existing cooking tools compared to blind people, highlighting the need for technology that considers their unique needs. The same study also identified key challenges LV people face, such as distinguishing objects with low contrast and safely interacting with dangerous kitchen tools. Wang \textit{et al.}~\cite{Wang2023} further interviewed six LV rehabilitation professionals to understand current training strategies and tools for cooking. They emphasized that existing solutions are insufficient to overcome all cooking challenges LV people experience. Our research fills this gap by creating an AR system that supports LV people in safely and efficiently using cooking tools through visual augmentations.




\subsection{Using Wearable AR to Enhance Accessibility}
In accessibility and HCI, wearable AR has been used to support people with diverse disabilities. For example, AR glasses can caption and visualize speech and sounds for deaf or hard of hearing (DHH) people~\cite{Findlater2019, Jain2018a, Jain2018b, Jain2015, Miller2017, Olwal2020, Peng2018, Schipper2017}, support hands-free interactions with screen displays for people with upper body motor impairments~\cite{McNaney2014, Malu2014, Malu2015}, offer speech support for people with aphasia~\cite{Williams2015}, and provide social cue therapy for children with autism~\cite{Washington2017}.

Within the low-vision aid context, head-worn AR devices can selectively enhance users' vision by interpreting their environment and tasks~\cite{Azenkot2017, Zhao2017}. For instance, prior research has developed AR systems that can capture real-time video feeds and apply image processing techniques~\cite{Massof1992, Zhao2016, Deemer2018} to enhance visual information, such as edge enhancement~\cite{Massof1994, Kwon2012, Hwang2014}, scene recoloring based on distance~\cite{Everingham1999, Hicks2013, Van2015}, and pixel remapping for visual field loss~\cite{Loshin1989, Gupta2018, Peli2001, Luo2006}. However, while these solutions are beneficial for simple tasks like reading~\cite{Sterns2017, Huang2019, Deemer2018}, they still lack semantic understanding of the scene and cannot effectively support more complex activities involving object interactions like cooking~\cite{Wang2023}.

More recently, researchers have combined AR and CV to create scene-aware visual augmentations to assist LV people with more intricate activities like visual search~\cite{Zhao2016}, stair navigation~\cite{Zhao2019}, wayfinding~\cite{Zhao2020}, obstacle maneuvering~\cite{Fox2023}, button pressing~\cite{Lang2021}, and sports~\cite{Lee2023}. Nonetheless, no prior systems have addressed the unique challenges of cooking involving dynamic tool interactions. Moreover, prior AR research for low vision primarily focuses on designing and evaluating visual augmentations. Therefore, they tend to oversimplify CV recognition in system development, such as by using QR codes~\cite{Zhao2016,Zhao2019} or existing spatial mapping APIs~\cite{Zhao2020,Fox2023} to anchor augmentations to the real world. This approach neglects the technical challenges of building a real-time AI-powered AR system and the potential impacts of technological limitations (\textit{e.g.}, CV errors, system latency) on user experience.

Our research advances the field by introducing CookAR, a wearable AR system that recognizes and augments kitchen tool affordances in near real-time, enhancing safe and efficient interactions during cooking.

\begin{figure*}[hbt!]
  \centering
  \includegraphics[width=\linewidth]{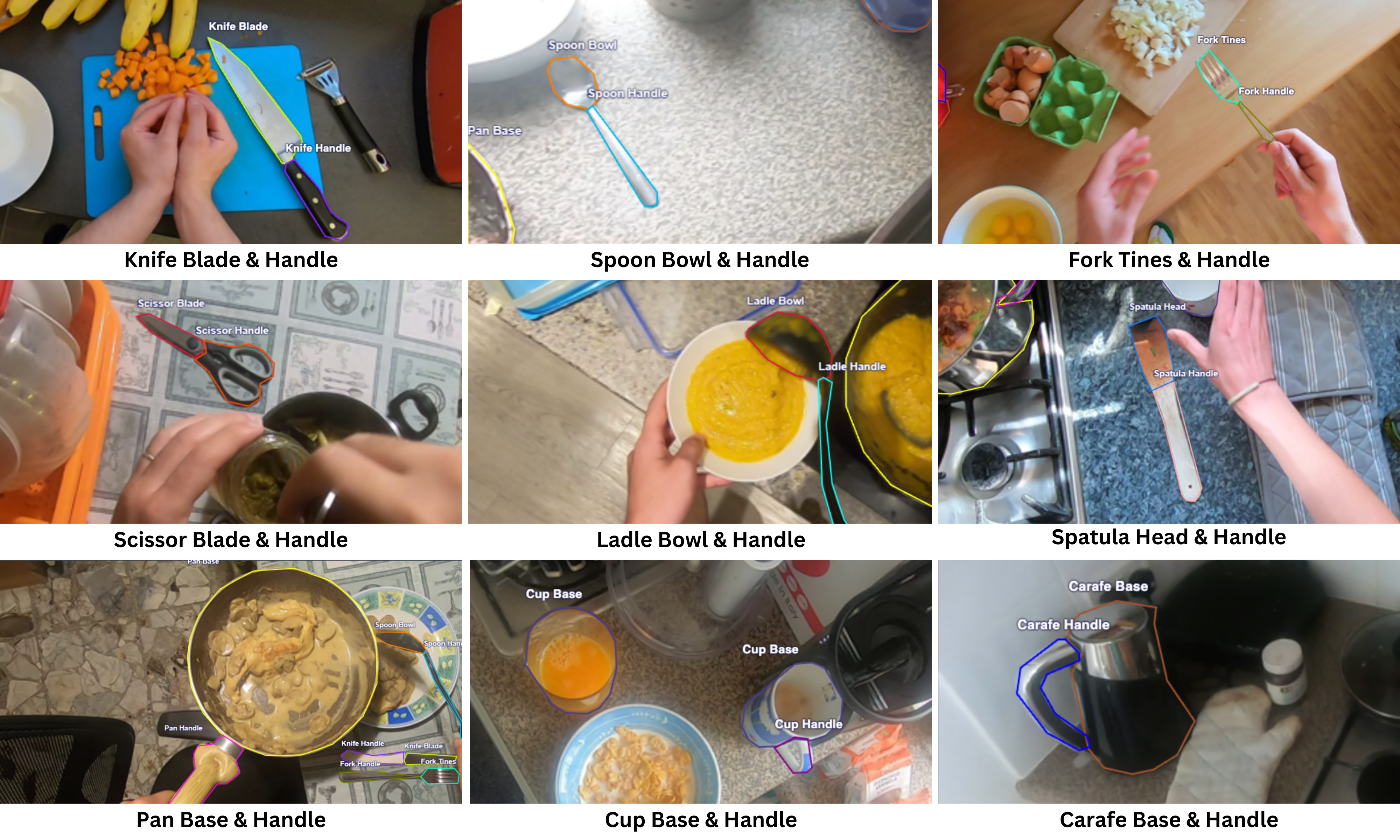}
  \caption{Example Roboflow annotations for each object class in our dataset (18 classes total).}
  \label{fig:annotations}
\end{figure*}

\subsection{Affordance Segmentation}
In contrast to most prior studies that augment objects as whole~\cite{Zhao2016, Fox2023, Lee2023}, our research focuses on recognizing and augmenting tool affordances. Affordance is traditionally defined as ``\textit{the opportunities for actions that objects offer, relative to the user's ability to perceive and act on them}''~\cite{Gibson1996, Gibson1997, Gibson2014}. Highlighting object affordances can effectively guide human attention and actions~\cite{Garrido2014, Tunnermann2015, Rehrig2022}. 
Despite the prominence of affordance segmentation in robotics~\cite{Chu2019a, Chu2019b, Nguyen2016, Nguyen2017, Bahl2023} and computer vision~\cite{Luddecke2017, Chuang2018, Luo2022}, automatic affordance augmentation has received comparatively less attention and applications in the field of HCI. Notably, there is a lack of an egocentric affordance dataset specifically created for the needs of LV individuals. To address this gap, we first created a new dataset focused on kitchen tool affordances by selecting and annotating image frames from the egocentric \textit{Epic Kitchens} video dataset~\cite{Damen2018} and fine-tuned an instance segmentation model. We then built an AR system that can segment and enhance tool affordance information.


\section{System Implementation}
To support safe and efficient LV hand-object interactions with kitchen tools, we designed and built \textit{CookAR}, a wearable stereo AR prototype that recognizes and augments cooking tool affordances in near real-time. Unlike traditional enhancements that target objects as a whole, our prototype highlights their affordances (\textit{i.e.,} functional parts), facilitating identification and interactions with areas to grasp or avoid. To create a fully-functional wearable AR system, we needed to address both the computer vision problem of accurately recognizing object affordances in real-time and the HCI problem of designing and rendering suitable affordance augmentations. In this section, we describe our approach for each step, including (1) collecting and annotating a dataset focused on the affordances of kitchen tools; (2) fine-tuning an instance segmentation and recognition model on our dataset to detect these affordances; and (3) building a head-mounted AR system with a stereo camera to render visual augmentations on the recognized tool affordances. \rev{The labeled dataset and fine-tuned model weight are one contribution of our work and are open-sourced at: \textit{\textcolor{blue}{\url{https://github.com/makeabilitylab/CookAR}}}}.

\subsection{Data Collection and Annotation}
To train a CV model for affordance segmentation, we first needed a labeled dataset. However, to our knowledge, there is no prior cooking tool dataset with annotations to enable affordance segmentation. Below, we describe our multi-step process to collect and annotate object affordances in egocentric cooking images.

\textbf{Data Collection.}
To build our kitchen tool affordance image dataset, we used an egocentric video repository called \textit{Epic Kitchens} \cite{Damen2018}, which consists of 100 hours of video footage of sighted people cooking in their homes. We selected this dataset since it not only involves a wide range of cooking scenarios with various kitchen tools but also captures video feeds from a first-person perspective, which aligns with the egocentric nature of head-worn AR devices. 

Because the Epic Kitchens dataset is large, we first needed to filter for frames of interest. We used \textit{YOLOv8}~\cite{Ultralytics2023} trained on the \textit{MS COCO} dataset~\cite{Lin2014} to detect and collect frames featuring cooking-related objects, such as spoons, knives, forks, cups, scissors, sinks, and dining tables. To minimize repetition, we skipped 20 frames after finding at least one of those objects. We then manually reviewed the selected frames to empirically remove similar, excessively blurry, or irrelevant images, resulting in 4,928 key frames. 

\textbf{Data Annotation \& Augmentation.}
We then labeled these frames using \textit{Roboflow}\footnote{https://roboflow.com}, an online tool for annotating, training, and optimizing CV models. Roboflow also supports labeling automation using the \textit{Segment Anything model} (SAM)~\cite{Kirillov2023}, allowing us to easily select and segment interactive parts of objects (\textit{e.g.,} knife blade \textit{vs.} knife handle) and add corresponding class labels.

\rev{Drawing on prior research~\cite{Li2021, Li2024, Wang2023}, the research team identified 18 distinct classes of kitchen tools} commonly used by BLV individuals: \textit{Knife Blade}, \textit{Knife Handle}, \textit{Spoon Bowl}, \textit{Spoon Handle}, \textit{Fork Tines}, \textit{Fork Handle}, \textit{Scissor Blade}, \textit{Scissor Handle}, \textit{Ladle Bowl}, \textit{Ladle Handle}, \textit{Spatula Head}, \textit{Spatula Handle}, \textit{Pan Base}, \textit{Pan Handle}, \textit{Cup Base}, \textit{Cup Handle}, \textit{Carafe Base}, and \textit{Carafe Handle} (Figure~\ref{fig:annotations}). When labeling, we adhered to the following heuristic: (1) the object should visually resemble the class it is labeled as; and (2) the object should serve functions similar to those of the label class. For instance, a large wooden spoon can be tricky to label, as it can resemble a spoon, ladle, or spatula, and have versatile use such as stirring a pan (like a spoon), scooping contents from a pot (like a ladle), or lifting eggs (like a spatula) across different images in the dataset. We labeled these ambiguous objects based on their shape and use in a given frame. \rev{Six research team members performed the annotations, each labeling a subset of images and having their work reviewed by another researcher to reduce errors and bias.}

After annotating, we used various image augmentation techniques available on Roboflow to enhance the dataset for better generalizability across real-world scenarios, including: cropping with 0\% minimum zoom and 40\% maximum zoom, rotation between -15° and +15°, brightness between -15\% and +15\%, blur up to 2.5px, and noise up to 0.1\% of pixels. We then adjusted the images to fit a 640x480 resolution (\textit{i.e.,} the MS COCO average image resolution~\cite{Lin2014}) to accommodate our chosen base model's preferences and facilitate their use in future research. This resulted in a final dataset of 10,152 images.

\begin{figure*}[hbt!]
  \centering
  \includegraphics[width=\linewidth]{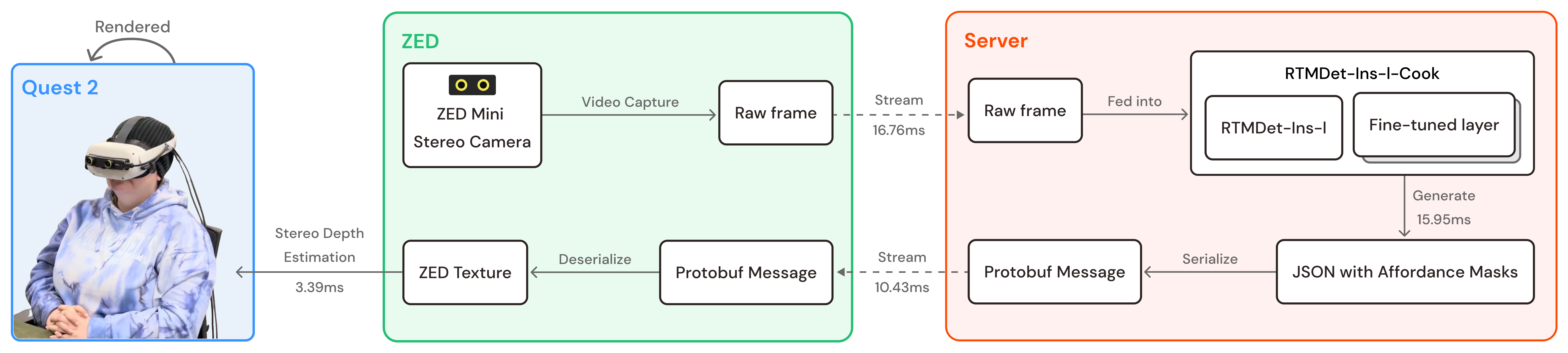}
  \caption{System overview of CookAR showing how data flows from the ZED Mini Stereo camera to an external camera for affordance segmentation, then sent back to ZED for rendering on the Quest 2 headset.}
  \label{fig:system_overview}
\end{figure*}

\subsection{Model Fine-Tuning}
To provide real-time object affordance information to LV users, we fine-tuned the \textit{RTMDet} model~\cite{Lyu2022}, specifically its \textit{RTMDet-Ins-l} variant, on our kitchen tool affordance dataset. This model is the current state-of-the-art in real-time instance segmentation\footnote{https://paperswithcode.com/sota/real-time-instance-segmentation-on-mscoco}, offering robust accuracy and 300+ FPS on an NVIDIA 3090 GPU. \rev{RTMDet features large kernel depthwise convolution and batch normalization layers, pre-trained on MS COCO~\cite{Lin2014}.}


We opted to fine-tune RTMDet instead of training it from scratch, as this allowed us to better leverage our smaller, class-specific dataset. \rev{To achieve this, we leveraged the fine-tuning pipeline provided by the \textit{MMDetection} library~\cite{Chen2019}, a PyTorch-based open-source toolkit for object detection and segmentation, which supports various models including RTMDet.} We initialized the base RTMDet-Ins-l model with pre-trained weights, froze its backbone, and adjusted the model configuration file for our label classes before training it on our kitchen tool affordance dataset. This customized model, dubbed \textit{RTMDet-Ins-l-Cook}, was trained over 150 epochs with a batch size of 4 on a single CUDA-enabled NVIDIA 4080 GPU. Because RTMDet-Ins-l-Cook underwent fine-tuning on a dataset with affordance annotations, it can mimic an affordance segmentation model's capabilities while retaining RTMDet's real-time performance. We provide a technical evaluation of our model in Section \ref{sec:tech-eval}. \rev{We also open-sourced our fine-tuning steps and code, giving researchers the tools to expand our dataset and fine-tune their own models: \textit{\textcolor{blue}{\url{https://github.com/makeabilitylab/mmdet-fine-tuning}}}.}

\subsection{The CookAR Prototype}
With our RTMDet-Ins-l-Cook model, we built CookAR, a wearable AR prototype that can recognize and visually augment the affordances of kitchen tools in near real-time. To implement CookAR, we addressed three key technical and HCI challenges, including: (1) how to spatially highlight object affordances in 3D space; (2) how to develop a real-time system pipeline to provide affordance feedback with minimal latency; and (3) how to best visually indicate affordances to LV users to support but not overwhelm their existing visual perceptions. See video figure for a demonstration.

To generate visual augmentations that align with the object parts in 3D space (\textit{e.g.,} a knife grip, a cup handle), \rev{we built a custom stereo video see-through AR system by combining the \textit{ZED Mini} stereo camera with an \textit{Oculus Quest 2} VR headset. While off-the-shelf AR headsets such as the Microsoft HoloLens 2 may eliminate the need for a bulky video passthrough system, they do not yet support long range real-time depth sensing.} We visualize the affordance representations as colored polygon overlays. Though the colors are settable, we currently use \colorbox[HTML]{2a2a2a}{\textcolor[HTML]{3BE8B0}{green}} (hexcode \#3BE8B0) to indicate a graspable area and \colorbox[HTML]{1a1a1a}{\textcolor[HTML]{FC626A}{red}} (\#FC626A) to indicate a risky area (Figure~\ref{fig:teaser}). As described in our user study section, we also further explored and brainstormed other affordances and augmentation designs.

Because our real-time CV model is computationally expensive, CookAR is tethered to a laptop with a NVIDIA 4080 mobile GPU. The CookAR system first captures image frames using the ZED Mini stereo camera and streams them to an external server via the \textit{Transmission Control Protocol} (TCP) for affordance segmentation by our RTMDet-Ins-l-Cook model (confidence threshold of 0.4). Then, the server converts the resulting JSON with affordance masks and labels into a \textit{Protocol Buffers} message\footnote{https://protobuf.dev} for efficient streaming. This message is then sent back for processing by the ZED Mini API~\cite{ZED-Unity}, which deserializes the message back into a JSON and creates a ZED-compatible texture (or colored overlay) for each affordance mask. Finally, the ZED Mini performs stereo depth estimation and overlays the textures onto the left and right image frames for binocular vision in the Quest 2 headset. \rev{To enable participants to move freely during the study session, we connected the CookAR system to a computer using long (16 feet) cables.}

\rev{In our latency analysis, we ran CookAR for five minutes and computed the average latency of each component: video streaming from ZED to the server took 16.76ms; affordance recognition took 15.95ms; result streaming back to ZED took 10.43ms; and depth estimation and augmentation rendering took 3.39ms. All other components had negligible impact on runtime. The overall latency is on average 46.82ms per frame ($\sim$21.36 FPS), resulting in a near real-time system. See Figure~\ref{fig:system_overview} for a system flow diagram.}


\begin{figure*}[hbt!]
  \centering
  \includegraphics[width=\linewidth]{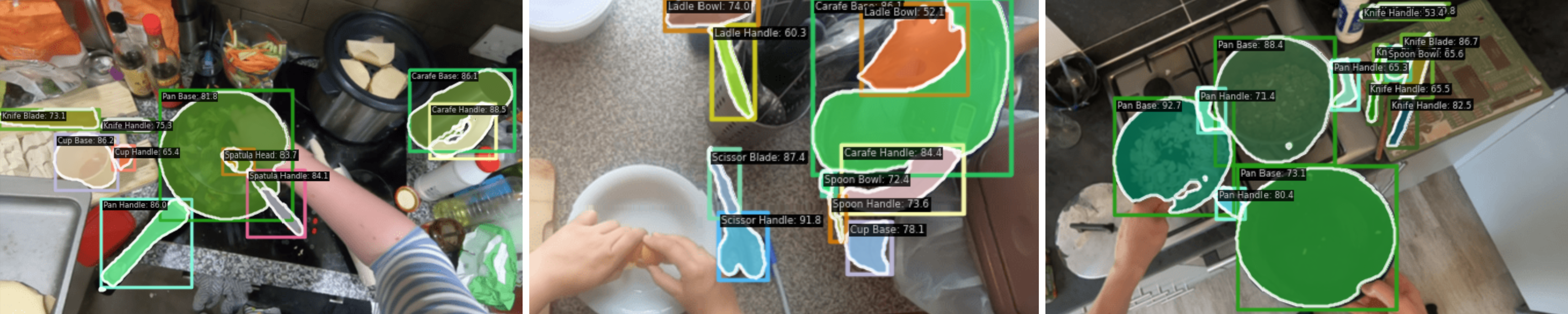}
  \caption{Example inferencing results on images from the test subset of our dataset. These images demonstrate how the RTMDet-Ins-l-Cook identifies and segments graspable, safe areas—even in the presence of hands or other partial occlusions.}
  \label{fig:inference}
\end{figure*}

\section{Technical Evaluation}
\label{sec:tech-eval}
We first conducted a technical evaluation of our RTMDet-Ins-l-Cook model, comparing its performance against the base RTMDet-Ins-l model on our kitchen tool affordance dataset. Findings indicate that our model is significantly more accurate in recognizing and segmenting affordances of cooking tools than the unmodified model.

\subsection{Methods}
To assess the performance of the base and fine-tuned models on our kitchen tool affordance dataset, we used MMDetection's~\cite{Chen2019} model testing pipeline, which performs evaluations using the test subset of a given dataset. With Roboflow, we generated a test set of 596 images with an 82-12-6 train-validation-test split and ensured that our model was not exposed to images in the test subset.


For instance segmentation tasks, accuracy is conventionally measured using three key metrics: segmentation mean average precision (mAP), AP at a 50\% Intersection over Union (IOU) threshold (AP@50), and AP at a 75\% IOU threshold (AP@75)~\cite{he2017mask}. \textit{IoU}, central to these metrics, quantitatively evaluates the overlap between predicted segmentation masks and the actual ground truth, serving as a direct measure of accuracy in spatial alignment. We explain each metric in detail below:

\begin{itemize}[leftmargin=.2in]
    \item \textit{mAP} offers a comprehensive assessment of a model’s precision across various detection thresholds by averaging precision at multiple recall levels for each class. It also aggregates results across a range of Intersection over Union (IoU) thresholds, from 0.5 to 0.95 in steps of 0.05, providing a holistic view of model performance across different degrees of overlap between the predicted masks and the ground truth;
    \item \textit{AP@50} uses a precision of segmentation at the IoU threshold of 50\%, a more lenient measure that considers a prediction correct if the overlap with ground truth is at least half;
    \item \textit{AP@75} evaluates precision at a stricter IoU threshold of 75\%, demanding higher accuracy in the overlap between the predicted segmentation and ground truth.
\end{itemize}

We applied these metrics to compare our model's performance against the baseline, aiming to capture the nuances of improvement across different levels of strictness in segmentation accuracy.


\begin{table}[t!]
\begin{tabular}{ l|ccc } 
\toprule
 Model Name & mAP & AP@50 & AP@75\\ 
\midrule
 RTMDet-Ins-l (on COCO) & 0.437 & 0.660 & 0.470\\ 
\midrule
 RTMDet-Ins-l (on our dataset) & 0.123 & 0.199  & 0.310\\ 
 RTMDet-Ins-l-Cook (on our dataset) & \bfseries0.463 & \bfseries0.749 & \bfseries0.486 \\ 
\bottomrule
\end{tabular}
\vspace{5px}
\caption{Affordance segmentation results. Our fine-tuned affordance segmentation model, \textit{RTMDet-Ins-l-Cook}, achieves superior performance across all metrics on our kitchen tool affordance dataset, outperforming the state-of-the-art baseline \textit{RTMDet-Ins-l} model. For reference, we also include \textit{RTMDet-Ins-l} results on the COCO dataset.}
\vspace{-7mm}
\label{tab:tech_eval}
\end{table}

\subsection{Results}
Our findings (Table~\ref{tab:tech_eval}) indicate that the base RTMDet-Ins-l model performs poorly on our affordance dataset despite its competency on the COCO dataset~\cite{Lin2014} with a 43.7\% segmentation mAP. Moreover, our fine-tuned RTMDet-Ins-l-Cook model excelled in identifying and segmenting cooking tool components, demonstrating a significantly higher segmentation mAP of 46.3\%, compared to the base model's 12.3\%. This improvement was also evident in our model's performance at different IoU thresholds, with AP@50 and AP@75 reaching 74.9\% and 48.6\%, significantly outperforming the base model's 19.9\% and 13.2\%. In Figure~\ref{fig:inference}, we show several example inference results of RTMDet-Ins-l-Cook on test images. Our model demonstrates impressive robustness, identifying and segmenting graspable, safe areas even when hands or other partial occlusions are present in the images. Overall, these findings highlight the enhanced accuracy of our RTMDet-Ins-l-Cook model.

\section{User Study} \label{sec:user-study}
As a complement to our technical evaluation, we conducted a three-part qualitative lab study with 10 low vision (LV) participants. Our goals were threefold: first, to evaluate how LV participants might benefit from real-time object affordance augmentations when completing cooking tasks; second, to solicit their reactions to a fully-functional but early-stage prototype (\textit{e.g.,} how do they react to augmentation errors); finally, to co-brainstorm visual overlay designs via design probes. Participants provided feedback throughout the study and answered open-ended questions regarding their experiences, which were recorded and transcribed for later analysis.


\rowcolors{2}{gray!8}{white}
\begin{table*}[hbt!]
\centering
\begin{tabular}{ccccc>{\raggedright\arraybackslash}p{8.4cm}}
\toprule
\textbf{P\#} & \textbf{Gender} & \textbf{Age} & \textbf{Left Eye Acuity} & \textbf{Right Eye Acuity} & \textbf{Description of Visual Field} \\
\midrule
P1 & Male & 30 & No Light Perception & 20/400 & Coloboma dominates the right superior portion of my right eye. \\
P2 & Female & 83 & 20/200 & 20/100 & Deteriorating eyesight from dry macular degeneration. Lost central vision on left eye. Central vision on the right eye is still there but not good. Have peripheral vision on both. \\
P3 & Female & 62 & 20/125 & 20/100 & Low vision. Some holes in it, like black spots. Scar tissue. \\
P4 & Male & 65 & 20/20 & 20/60 & Can see from 2/3’s of left eye, some far right peripheral vision from right eye. \\
P5 & Female & 70 & 20/200 & 20/100 & Macular degeneration and side effects of chemotherapy. Blurry vision and need font enlargement to read. Visual field intact. \\
P6 & Male & 50 & 20/40 & No Light Perception & Blind in right eye. Need glasses for left. Visual field intact. \\
P7 & Female & 81 & 20/200 & 20/60 & Diminished vision due to macular degeneration. Visual field intact. \\
P8 & Female & 80 & 20/50 & 20/50 & Have dry macular degeneration with loss of some vision in the center of my left eye. \\
P9 & Female & 30 & Light Perception & 20/80 & Can make out faces w/right eye. Left eye blind. Visual field intact. \\
P10 & Female & 71 & 20/60 & 20/100 & I have Glaucoma. My field of vision is 5\% eyesight. 5\% in my left and 5\% in the right remaining. \\
\bottomrule
\end{tabular}
\vspace{3pt}
\caption{Individual study participant information, including their gender, age, left and right eye acuity, and a self-reported description of their vision.}
\vspace{-20pt}
\label{tab:participants}
\end{table*}

\subsection{Participants}
To achieve a diverse participant pool, we recruited 10 LV participants from two cities (Madison, WI and Seattle, WA) via mailing lists and snowball sampling. Participants were screened using a demographic questionnaire, which gathered information on age, gender, vision condition, and prior experience with AR and AI technologies. The average age was 62.2 years (\textit{SD=}19.6), with a gender distribution of 70\% female and 30\% male. Participants had a broad range of low vision conditions with visual acuity ranging from 20/40 to 20/400 and visual field loss at different areas---see Table~\ref{tab:participants}. Most participants reported little to no experience with AR and AI, except for P1 who had used both technologies.

\subsection{Apparatus} \label{sec:apparatus}
The study was conducted in a well-lit lab environment. Participants sat in front of a large table, where we placed nine different kitchen tools---knife, spoon, fork, scissors, ladle, spatula, pan/pot, cup, and carafe. We used a dark green table cloth to simulate a visually challenging environment with low contrast. We also prepared a yellow wooden cutting board, a bowl, a piece of cheese, and a stick of butter for the participants to use in the study, although CookAR can only recognize and augment the nine aforementioned kitchen tools at the current stage. Lastly, we recorded the experiment using a laptop and a smartphone on a tripod.

\subsection{Procedure}
The single-session 90-minute study consisted of three phases. In Part 1, we asked participants to grab cooking utensils \rev{with CookAR and two baselines.} In Part 2, participants completed a full cooking task where they made macaroni and cheese while using CookAR with affordance augmentations. Finally, in Part 3, participants brainstormed tool affordances and desired augmentation designs while examining and critiquing design probes. Prior to the study tasks, participants completed a tutorial to become familiar with CookAR. \rev{In total, participants interacted with our AR device for $\sim$30 minutes across the study: a 5-minute tutorial, 10 minutes of object grabbing, and 15 minutes of free-form cooking.} We provide more details below. \rev{The full study protocol is in the Appendix.}

\textbf{Tutorial.} Participants first completed a tutorial task, where they interacted with a cooking pot using CookAR. Participants wore and adjusted the Oculus Quest 2 headset and freely explored CookAR and its affordance augmentations. Once they achieved a comfortable fit and an understanding of CookAR, the study proceeded.

\begin{figure*}[hbt!]
  \centering
  \includegraphics[width=\linewidth]{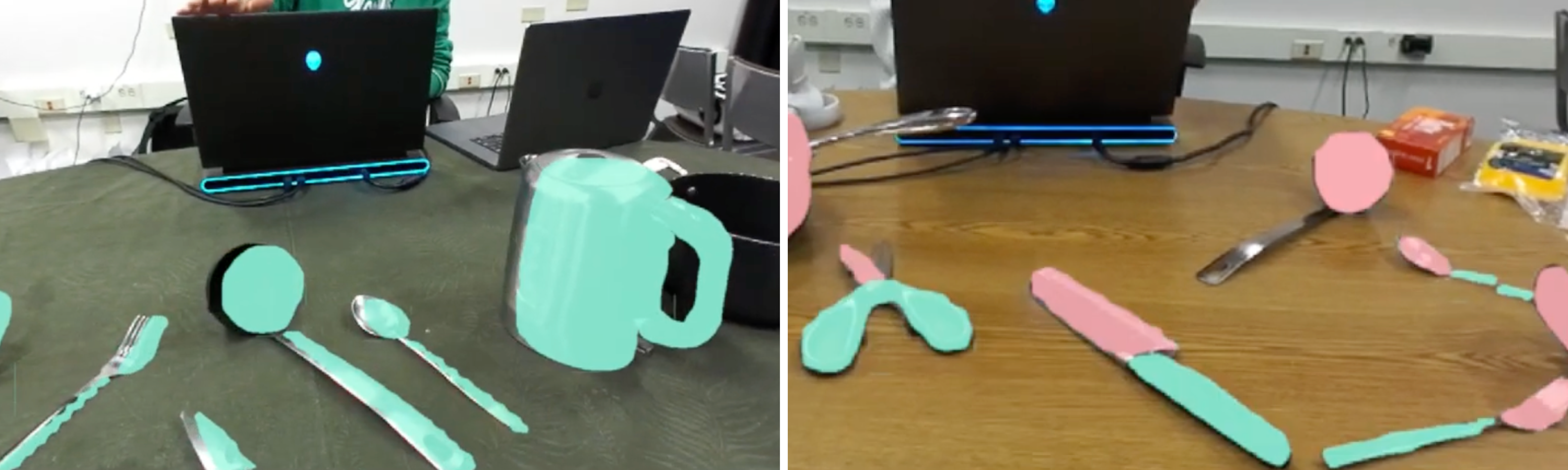}
  \caption{The CookAR prototype with whole object augmentations (left) and affordance augmentations (right). The whole object augmentations are green instance segmentation masks, while the affordance augmentations are green (grabbable) and red (hazard) affordance segmentation masks.}
  \label{fig:conditions}
\end{figure*}

\textbf{Part 1: Tool Grabbing Task.} In Part 1, participants were asked to locate and pick up cooking tools under \rev{three conditions: (1) real-world baseline (\textit{i.e.,} without CookAR), (2) augmentation baseline (\textit{i.e.,} whole object augmentations)~\cite{Zhao2016,Fox2023}, and (3) CookAR (\textit{i.e.,} affordance augmentations)} (See Figure~\ref{fig:conditions}). We counterbalanced the condition order via \textit{Latin Square}. In each condition, participants conducted five trials of picking-up tool tasks. We randomly chose a cooking tool per trial from the nine kitchen tools on the experiment table (See Section \ref{sec:apparatus}). Participants were asked to keep their eyes closed until the researcher named an object to reduce the effect of memory on task performance. Moreover, the researcher rearranged the placement and angle of the cooking tools between each condition. After each condition, we asked participants three 7-point Likert questions about effectiveness, comfort, and distraction, as well as open-ended questions regarding their experience with each augmentation condition. After all 15 trials, we asked participants to compare the pros and cons of the three conditions, suggest improvements for CookAR, and identify potential applications of CookAR outside of kitchen contexts.

\begin{figure*}[hbt!]
  \centering
  \includegraphics[width=\linewidth]{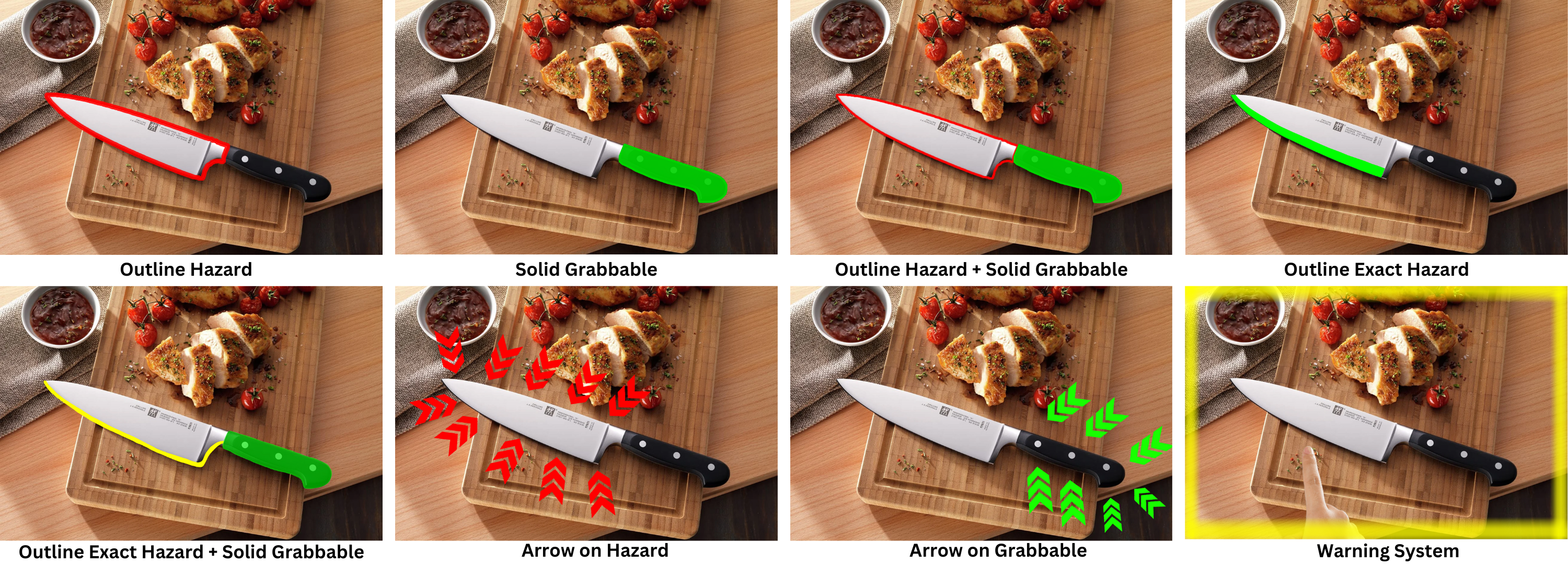}
  \caption{Design probes used in Part 3 of the study to spark design ideas.}
  \label{fig:probes}
\end{figure*}

\textbf{Part 2: Full Cooking Task.} In Part 2, we asked participants to cook a macaroni and cheese dish using CookAR with affordance augmentations. We provided participants with step-by-step instructions for consistency and to ensure participants interacted with all nine objects CookAR can recognize: (1) grab a cup of water and pour it into a \underline{carafe}, (2) pour water into a \underline{pot} using a \underline{carafe}, (3) cut a piece of butter using a \underline{knife}, (4) cut a piece of cheese using a pair of \underline{scissors}, (5) put the macaroni, butter, and cheese into a \underline{pot}, (6) stir with a \underline{spoon}, (7) stir with a \underline{spatula}, (8) place the finished macaroni and cheese in a bowl using a \underline{ladle}, and (9) pick up a \underline{fork} and enjoy. For the safety of our participants, we supplied a knife with a dull edge and avoided the use of heat. As participants completed this task, they were encouraged to think aloud, articulating how the affordance augmentations supported or hindered their activities, how CookAR impacted their overall cooking experience, and any suggestions they had for augmentation designs. After completing the free-form cooking task, we asked participants to reflect on these same topics through open-ended questions.

\textbf{Part 3: Brainstorming and Co-Design.} In Part 3, we asked participants to brainstorm future designs and applications of CookAR. Drawing on prior work in low vision augmentation~\cite{Zhao2016, Zhao2019, Fox2023, Lang2021}, we created and presented design probes of various augmentation designs (Figure~\ref{fig:probes}) and asked participants for feedback. The design probes included: (1) outlines to reduce visual clutter in comparison to solid-colored overlays, (2) solely displaying either the grabbable or the hazardous augmentation, (3) highlighting the more specific hazardous part such as the sharp edge of a knife blade rather than the whole blade, (4) employing arrows to widen the area covered by the augmentations, and (5) introducing a visual warning system when the user's hand gets too close to a risky area. After presenting our own designs, we invited participants to propose other ideas for both simple objects like knives as well as more complex objects with additional interactive parts beyond grabbable and hazardous areas, like a carafe, which has many openings and buttons with different purposes. Lastly, we asked participants to identify other scenarios where affordance augmentations might be beneficial.

\subsection{Analysis}
We recorded participants' quotes using Zoom. Transcriptions were first done by the video conferencing software, then the research team manually revised the transcripts. We collected 346 distinct quotes across our 10 LV participants, which we analyzed using reflexive thematic analysis~\cite{Braun2006, Braun2019}. The first author, who facilitated every user study session, created an initial codebook by reviewing the revised study transcripts. The research team then collaboratively iterated on the codebook while checking for bias and coverage. With a final codebook consisting of 23 codes, the first author coded participants' quotes, after which the team discussed the resulting themes. For Likert score analysis, we used a Wilcoxon signed-rank test since the data does not follow a normal distribution.

\section{Results}
In our three-part qualitative study, participants completed tool-grasping tasks, a free-form cooking task, and a brainstorming session with design probes. Overall, participants found the real-time affordance augmentations helpful when interacting with various cooking tools. They also suggested desired augmentation designs and key affordance parts. We expand on these findings below.

\subsection{Affordance vs. Whole Object Augmentations}
\label{sec:comparison}
All but one participant (P6) preferred affordance augmentations over whole object augmentations for supporting kitchen tool interactions. They noted a trade-off between the augmentations' utility and distraction, with the former generally outweighing the latter: ``\textit{Seeing one color was less distracting than seeing two colors. But you’d have to know which end of the tool is the handle and which is the working end}'' (P2). We report participants' feedback on the effectiveness and distraction of the augmentations below.

\textbf{Effectiveness.} In examining Likert scores on perceived effectiveness, we found no significant difference between whole object augmentations and affordance augmentations ($W\!=\!36.5$; $p\!=\!0.32$). However, participants on average gave higher ratings to affordance augmentations ($mean_a\!=\!5.3$; $SD_a\!=\!1.6$) over whole object augmentations ($mean_w\!=\!4.6$; $SD_w\!=\!1.4$). Affordance augmentations are advantageous in quickly understanding the overall scene (P1, P10), along with the placement and orientation of individual objects: ``\textit{It helps to have two colors. I could see that more readily and quickly to understand how to use the object and how it is placed... your system is helpful because where the tool starts and ends and where the handle starts and ends is more clear}'' (P5). In addition, affordance augmentations become particularly useful when handling objects that have hazardous (9 out of 10 participants) (\textit{e.g.,} sharp, hot ) or small (8/10) (\textit{e.g.,} door handles, buttons on appliances) parts, or have insufficient color contrast (7/10) (\textit{e.g.,} all silver or black cooking tools): ``\textit{You've gotta show parts you can and shouldn't grab. Green tells me that's a safe place to go with my hand. Anything not green, I shouldn’t grab... It can help me avoid dangerous parts or perhaps even find small things like remote controllers}'' (P4). Furthermore, four participants expressed that for objects with more complex interaction components than ``\textit{just grab and don't grab}'' (P7), like a carafe with its handle, base, buttons, lid, and spout, they would accept the use of more than two colors, although ``\textit{more than four colors can be quite distracting}'' (P7). We discuss additional augmentation designs in Section~\ref{sec:augmentation_design}.

\textbf{Distraction \& Comfort.} While most participants qualitatively expressed that the whole object augmentations are less distracting than the affordance augmentations (6/10), the comfort and distraction Likert scores were not significantly different ($W\!=\!52.5$; $p\!=\!0.88$ and $W\!=\!46$; $p\!=\!0.78$ respectively). The difference in average rating was also negligible, although participants on average found whole object augmentations to be slightly more comfortable ($mean_w\!=\!5.1$; $SD_w\!=\!1.3$ vs. $mean_a\!=\!5.0$; $SD_a\!=\!1.2$) and less distracting ($mean_w\!=\!2.3$; $SD_w\!=\!0.9$ vs. $mean_a\!=\!2.5$; $SD_a\!=\!1.1$). P6, who preferred whole object augmentations, said ``\textit{I think the more colors you have, the more distracting it becomes. So I prefer just the whole object in green than having 2 or 3 different colors. An outline would be better. I would definitely stay away from multicolored and just stick with one color. I can figure out its different parts.}'' Additionally, three participants shared that the whole object augmentations could be more useful depending on the scenarios. For tasks such as locating or avoiding objects, where interaction is not the goal, whole object augmentations are more preferable, since they are less distracting: ``\textit{If I am looking for the remote controller, if it could make the remote stand more out in green or something. I don't need its parts}'' (P3).

In summary, we found that if a person intends to interact with an object, affordance augmentations are more helpful than whole object augmentations. Conversely, in cases where interaction is not the objective, whole object augmentations may be preferred as they are less distracting. \rev{Despite the positive qualitative feedback on CookAR, the Likert scores for the real-world baseline (without CookAR) were higher than those for both the whole object augmentations (effectiveness: $W\!=\!94$; $p\!<\!.001$, comfort: $W\!=\!95$; $p\!<\!.001$, $W\!=\!10$; distraction: $p\!<\!.001$) and affordance augmentations (effectiveness: $W\!=\!77.5$; $p\!<\!.05$, comfort: $W\!=\!52.5$; $p\!<\!.001$, distraction: $W\!=\!10$; $p\!<\!.001$). This may be due to current limitations of AI-powered AR systems such as accuracy and latency. We discuss this further in Section \ref{sec:challenge}}.

\subsection{Free-form Cooking with CookAR}
\rev{All participants were able to complete all free-form cooking steps within three to five minutes.} However, due to technical limitations in accurately segmenting affordances, deploying heavy CV models, and rendering spatially accurate overlays in real-time on AR headsets, participants experienced recognition errors and latency with CookAR. All participants observed ``\textit{flickering}'' and inaccurate augmentations. Participants also pointed out that ``\textit{the colors took some time to catch up}'' (P3) as they quickly rotated their head.

Nonetheless, all but P6 saw potential in CookAR to assist with kitchen tool interactions and beyond: ``\textit{I like the contrasting color. I just wish it more closely matched the object’s actual location. I think this highlighting scene is a great start. If the system is perfect, the dual color highlight system would be great and most useful. The system would be perfect if I am in a kitchen or just trying to grab really anything}'' (P1). P1, P5, P9, and P10 were particularly excited as they were able to better visually perceive object information: ``\textit{This is fun! I can also use my eyes more to see shape and how [a tool] can be used. I want to try your system again once you make it better}'' (P9). Participants also identified the following additional use cases for a CookAR-like system: cleaning (P3, P6, P7, P9), woodworking (P4, P5, P9), walking outdoors (P2, P6, P9), driving (P4, P6, P7), visiting a foreign kitchen (P2, P5), restaurants (P3, P9), gardening (P5, P10), watching sports (P7, P10), playing board games (P7, P10), going down stairs (P7, P9), identifying pill bottles (P7), and interacting with appliances with multiple buttons like a toaster (P1). 


\subsection{Desired Augmentation Designs} 
\label{sec:augmentation_design}
We report participants' preferences on augmentation designs for grabbable and hazardous areas based on the design probes.

\textbf{Combining solid and outline augmentations.} 
As opposed to solid-colored overlays, nine participants preferred a mix of solid and outline augmentations because solid colors are more salient, whereas outlines are less distracting: ``\textit{Solid colors are helpful because they grab my attention... outlines are helpful because I can still see the part I'm trying to use with less distraction}'' (P5). Among those nine participants, all but P7 preferred solid-colored overlays for the grabbable area and outlines for the hazardous area because ``\textit{the grabbable area is the most important}'' (P3, P4, P5, P8, P9), ``\textit{all you need to know is its shape}'' (P4, P9), and ``\textit{other parts should be outlined since you may want to do more with it, and solid color just makes it harder to use it}'' (P4, P8, P9). However, P7 preferred the outline for the grabbable area since it is less distracting and still shows the shape of the handle.


For the risky area to outline, P8 preferred highlighting solely the exact hazard (\textit{e.g.,} the sharp edge of the knife blade), as opposed to the entire dangerous part of an object (\textit{e.g.,} the whole knife blade), because she needs to know the relatively safer area for interaction. For example, as she described, ``\textit{I might grab the top of the blade when I’m dicing or chopping. This tells you exactly where you shouldn't touch.}'' In contrast, all other participants wanted the outline augmentation because it is less distracting, yet still defines the overall shape of the hazardous area (P4, P5, P7, P9, P10) and what it is used for (P4, P7, P9). For example, P9 said, ``\textit{I prefer to see the outlines on the [whole] blade, just so that way, you know which type of blade you’re grabbing. Cause a bread knife would look different from your knife. Some are thinner, some are fatter. People can be quite picky about their knife choices.}''

Lastly, P1, P4, and P10 expressed concerns that overlaying perfectly aligned solid-colored affordance augmentations can be technically challenging. They suggested a colored circle may be enough, since they only need ``\textit{a hint to see a glimmer of the object}'' and determine how the objects are oriented (P1).

\textbf{Enhancing color contrast.}
Using colors to distinguish object affordances was well-received, as participants often color code their own cooking tools: ``\textit{So I always try to get things color coded... especially if things are in drawers, it takes a lot of cognition for me to tell you what’s what. If it’s colored, it’s so much easier. This system is huge cause it's doing color coding for me}'' (P1).

Every participant favored using green for safe-to-grab and red for dangerous areas, as ``\textit{green signals `yes' while `red' signals no}'' (P4). However, P4, P7, and P9 struggled to clearly see our choice of red and requested a brighter shade of red, with P4 even suggesting white. Moreover, participants noted that the color contrast between the tool and the background is more important than the specific colors used, since many kitchen objects are white, silver, or black with low contrast. For example, when cooking mac and cheese in the study, most (8/10) participants found it challenging to cut butter and understand where the yellow butter starts and ends because it was on a yellow wooden cutting board. To address this, P3, P4, P5, and P10 suggested the system should automatically select colors that contrast against the background: ``\textit{The background you have it against will make a big difference, right? So on a darker background, I should be getting light colors}'' (P4). P7 jokingly said, ``\textit{I mean, a green stick of butter could be weird, but it would at least let me cut better}.''

\textbf{Auditory feedback.}
Instead of visual augmentations, all participants preferred auditory feedback for warning in urgent scenarios (\textit{e.g.,} when the user's hands get too close to a knife blade), as a visual warning could be easily missed by low vision users (P3, P5, P7, P9) and also makes the overall visual field busier (P4, P10). P3, P5, and P6 suggested short yet noticeable audio such as ``\textit{beep beep},'' while P7 and P10 preferred explicit verbal warning (e.g., ``\textit{stop}'') since small noises can also be generated by other devices, such as a microwave or a fridge. P4 and P9 further suggested the system should employ different auditory signals for different hazards.

\textbf{Action-aware augmentations.}
As opposed to constantly augmenting all affordances, half of the participants suggested generating augmentations based on users' current tasks or behaviors to reduce potential distraction. For example, with a knife, both the handle and blade can be augmented to start, then when a person grabs it, the handle augmentations could be turned off (P7); or, as a person gets close to a carafe with a cup of water, the rim of the carafe could be highlighted (P4). Moreover, seven out of 10 participants also suggested using voice commands to control the augmentations, such as turning on and off an augmentation or adjusting the augmentation design (\textit{e.g.,} colors or forms).

\subsection{Additional Tool Affordances} 
In addition to the grabbable and hazardous affordances focused on by our CookAR system, participants collectively suggested five other important affordances for kitchen tools: (1) entry area, (2) exit area, (3) containment area, (4) intersection area, and (5) activation area. We elaborate on these seven affordances along with participants' preferred augmentation designs.

\textbf{Grabbable area.}
A grabbable area is the part designed for safe handling or manipulation. This can include handles, grips, or any part intended for direct hand contact. For grabbable part of an object, participants preferred green solid-colored augmentations.

\textbf{Hazardous area.}
A hazard area is the part that poses potential risks or dangers to the user. This could include sharp edges, hot surfaces, or any part that can cause injury if touched or mishandled. For hazardous part of an object, participants preferred red outline augmentations.

\textbf{Entry area.}
An entry area is the part designed for initiating access, such as pouring. This could be the rim of a cup or a pot, the opening of a carafe, or any designated point that allows entry into an object’s containment space. All participants consistently noted that this area should be augmented by an outline rather than a solid color, as the latter obstructs relevant actions like pouring or scooping: ``\textit{The color blobs hide the item that you’re trying to put things into virtually completely. And so I can’t really tell if I am pouring something in correctly}'' (P4). 

\textbf{Exit area.}
An exit area is defined as the point through which contents are meant to be released. This could be the spouts, holes, or any defined pathway that guides content out of the object's containment space, and it can be the same as the entry area for some objects, such as bowls and cups. Several participants (4/10) suggested that the carafe's spout, similar to the entry area, should be outlined: ``\textit{Highlighting the spout would be helpful if you had to pour, because if I poured in the wrong place, I wouldn’t know until something spills. I think I can pour more effectively if you highlighted this by aligning it with the edge of a cup or something. An outline would be great so I can see the water flowing out}'' (P7).

\textbf{Containment area.}
A containment area has some depth and is meant to hold content within, such as food and liquid. This could be the interior of a cup or pot, the base of a spoon or ladle, or any defined space within the object that is meant to keep something in. The current solid-colored overlays in CookAR interfere with visibility of the containment space. Instead, all participants wanted CookAR to augment only the entry and exit areas using outlines, leaving the containment space without any augmentation.

Additionally, eight participants expressed that they need assistance with understanding the depth of the containment space and the amount of content it already holds. As P8 expressed, ``\textit{A lot of people with low vision cannot see inside and know how much water they can pour. So somehow showing the water level and size of the teapot [is helpful]. Mine is a lot bigger, it makes 12 cups or something, and it’s all black, so it’s even harder to see what’s inside.}'' While P9 has a strategy to overcome this challenge by using her finger to feel the liquid level, she cannot use it when the water is hot. She thus suggested the system rendering ``\textit{a blue disk}'' to indicate the water level. In terms of augmenting the depth of the containment area, participants suggested using a virtual line from the rim to the bottom of the pot (P3, P4, P5, P9, P10), a measuring tape with ticks (P4, P5, P9, P10), or a line with changing colors (\textit{e.g.,} a green line that turns red as water fills up), (P4, P5). P10 further suggested an auditory cue (\textit{e.g.,} a `ding' sound) to indicate action milestones, such as when water reaches quarter of a cup.

\textbf{Intersection area.}
An intersection area is where parts of two or more objects meet. This could be where a knife blade touches the butter for cutting or where a cup touches a pot for pouring. Interactions that require precise alignment between two objects are particularly challenging to our LV participants. Half of the participants suggested generating augmentations to highlight the intersections or relationships between two interacting objects, for example, the location where a knife cuts the butter (P5) or the alignment between a ladle and a bowl when pouring (P9). As P9 mentioned, ``\textit{Using a ladle has always been a problem for me. Pouring the ladle into things is usually the hardest part, because you never know if the ladle is in the right spot or too wide out of the way. Maybe, if you have the ladle on top of a bowl, [CookAR should render] a [virtual] shadow that gets casted onto the bowl.}''

\textbf{Activation area.}
An activation area is designed for initiating, activating, or turning on an object's function or features. This could be buttons, switches, touch-sensitive surfaces, or any interactive components that trigger the operation of an object. Participants identified activation areas on many household appliances, such as buttons and dials on stove tops, microwaves, or coffee pots. They are used for various purposes including starting a machine, opening a lid, and adjusting settings. For example, P8 said: ``\textit{I just bought a vacuum with multiple buttons. You would want different colors for the handles and buttons}'' Seven participants preferred outline augmentations for the activation area. Additionally, P10 further suggested a clock-like augmentation in addition to an outline for turnable dials: ``\textit{On a stove, I don't know what is medium heat. As I turn the knobs on a stove, the system could show me `2 o'clock,' `3 o'clock,' and so on. `6 o' clock' is probably a medium heat. `9 o'clock' is probably a high heat.}''

\begin{figure*}[hbt!]
  \centering
  \includegraphics[width=\linewidth]{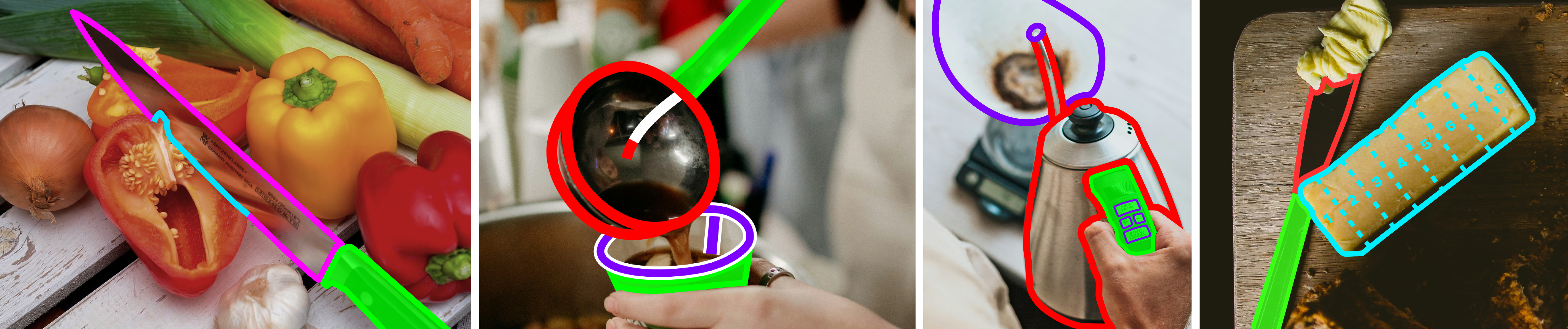}
  \caption{Envisioned augmentations based on participants' desired designs, including (1) highlighting not just the grabbable and hazard areas of a knife (using a color pairing that is not green and red for increased contrast) but also the contact location with the target object (\textit{e.g.,} a bell pepper); (2) illuminating the exit and entry areas of a ladle and cup along with their containment areas and indicating the depth and filled-up level of the containment area; (3) showing the activation areas (\textit{i.e.,} buttons) on a carafe along with the exit location of the spout; (4) showing the intersection of a butter knife with butter as well as measurement highlights overlaid on a stick of butter (\textit{e.g.,} tablespoons).}
  \label{fig:desired_designs}
\end{figure*}

\section{Discussion}
CookAR explores the use of real-time affordance augmentations to enhance kitchen tool interactions for LV people and advances the state-of-the-art in AI-powered AR systems. Results from our user study indicates a preference for affordance augmentations over whole object augmentations during tool interactions. Additionally, participants favored augmentation designs that incorporate both solid-colored and outlined overlays with contrasting colors. 
We discuss design implications for affordance augmentations as well as current limitations and future opportunities of AI-powered AR systems for low vision.

\subsection{Design Implications for Affordances}
Throughout the study, LV participants proposed a range of affordances for kitchen tools and indicated preferred augmentation designs. We summarize and expand on these suggestions.

\textbf{When to use affordance augmentations?} Our study findings suggest that visual augmentations should maximize utility and minimize visual clutter and confusion. As such, it is critical to render augmentations tailored to users' intent and reduce distraction. For example, affordance augmentations that involve multiple pieces and colors are more preferred to support direct hand-object interactions, while whole object augmentations are more suitable in general visual perception tasks such as avoiding obstacles and locating an object. Beyond a kitchen, affordance augmentations could also be applied to other scenarios, as our participants suggested, such as gardening, playing board games, and interacting with appliances (reaffirming Lang \textit{et al.}~\cite{Lang2021}).


\textbf{Where to apply affordance augmentations?} Affordances can refer to any object parts that indicate diverse actions or interaction opportunities. However, LV people face distinct interaction challenges, resulting in unique affordance opportunities. In our qualitative study, we identified seven essential affordances of kitchen tools that encapsulate important yet challenging interaction tasks for LV users. They include: (1) grabbable area, affording touching and handling action; (2) hazardous area, affording risks and avoidance; (3) entry, affording a target to aim at or pour in; (4) exit, affording pouring out and usually requiring accurate alignment with the entry of another object (\textit{e.g.,} food transferring or pouring); (5) containment area, affording holding content in and preferring augmentations on the content amount (\textit{e.g.,} ingredient measurement); (6) intersection area, affording touching or interaction between two objects; and (7) activation area, affording control features on an object. This affordance taxonomy summarizes the critical areas on objects as well as the hand-object (\textit{e.g.}, grabbable \textit{vs.} hazardous areas) and object-object (\textit{e.g.}, entry-exit alignment, intersection between objects) relationships during interactions. 

\textbf{How to augment affordances?} Different augmentations should be designed for different affordances according to the interaction tasks they indicate. In our study, participants preferred solid-colored overlays for grabbable areas to enable fast perception and action, measuring augmentations (\textit{e.g.,} line with ticks) for containment area to indicate content amount, and outlines for other affordances to avoid distraction and occlusion. In terms of colors, augmentations should adopt colors with high contrast against the environment. We also suggest leveraging cultural and semantic meanings of colors, such as green for safe-to-grab areas and red for risky areas. However, while preferred by the LV participants in our study, the green-red combination should be used cautiously given the prominence of red-green color vision deficiency.


\textbf{How to control affordance augmentations?} Due to the diverse visual abilities and preferences of LV users~\cite{zhao2015foresee}, future systems should support extensive personalization capabilities such as voice-based control for customization and automatic adaptations. For instance, users should be able to adjust different aspects of an augmentation, such as switching it on or off, choosing between solid-colored and outlined overlays, changing the outline thickness, and selecting suitable colors. Additionally, these systems should intelligently adapt by recognizing user actions or tasks to only highlight necessary object parts and signal warnings. They should also automatically alter the augmentation colors to generate high contrast against the background.

\rev{\textbf{Example designs.} Reflecting on these key design insights for affordance augmentations, we created some initial design mockups shown in Figure~\ref{fig:desired_designs}. Starting with solid-colored overlays for grabbable areas, measuring lines for containment areas, and outlines for all other areas, CookAR should allow customization, such as toggling overlays, adjusting outline thickness, and changing colors. Future work should explore these and other designs empirically.}

\subsection{Challenges in AI-powered AR Development}
\label{sec:challenge}
This paper presents several key technical contributions across CV and HCI by constructing the first egocentric kitchen tool affordance dataset, fine-tuning an affordance segmentation model on our dataset, and developing a fully-functional stereo AR system that generates real-time affordance augmentations. However, our study also revealed the impact of technology limitations on user experiences. For example, while finding CookAR promising, participants gave significantly higher Likert scores for the real-world baseline condition (\textit{i.e.}, without CookAR). Below, we reflect on key technical challenges stemming from both fields.


\textbf{AI models for real-world use.} Although our fine-tuned model outperforms the base model, its mAP is still too low to successfully support dynamic activities like cooking in real-world contexts. For instance, its AP@75 is 48.6\%, meaning in the worst case, about half of all predictions fail to achieve greater than 75\% overlap with the ground truth affordance masks, resulting in misalignment between augmentations and the original objects. The recognition results could become worse during real-world use on AR glasses due to natural human behaviors like users' constant head motions. For example, LV users tend to get much closer to view objects than sighted users~\cite{reynolds2024salient}. This issue highlights the evaluation gap between HCI and AI: a model that performs well under AI metrics (\textit{e.g.}, mAP, AP@50, AP@75) may be suitable for in-lab user studies but less so in naturalistic settings. We suggest that when developing AI models, researchers should consider the potential real-world use cases, human needs, and integration to different hardware platforms (\textit{e.g.}, wearable AR) to enable use in practice.

\textbf{Affordance models and datasets.} As opposed to object recognition models and datasets that attract significant attention in AI~\cite{Lin2014,deng2009imagenet,he2016deep,liu2021swin}, research on affordance models and datasets remains nascent. To address this issue, we collected and labeled an affordance image dataset for kitchen tools and fine-tuned an object detection model on the affordance dataset to balance accuracy and speed. However, due to the relatively small scale of the dataset and the RTMDet model not being designed for affordance, our system encountered affordance-specific issues. For example, our model often struggled to distinguish different handles, as many handles across various cooking tools look similar. While not affecting the mask generation (allowing users to still see the correct augmentations), it interfered with the object tracking model supported by the ZED Mini API, leading to flickering and unstable augmentation rendering. This is also noticeable in the video supplement.

To enhance system robustness and affordance recognition capabilities, future AI research should consider the following key areas: (1) Developing larger and more diverse affordance datasets. These datasets should capture a wider variety of object interactions and functionalities, allowing the model to learn from a richer set of scenarios; (2) Designing models for affordance detection, such as incorporating training objectives that refine object part relationship understanding for better affordance prediction; and (3) Improving object tracking algorithms to ensure more stable augmentation rendering, especially in dynamic environments where precise object localization is crucial. 

\textbf{System latency.} Latency is always a concern for AR systems, especially since off-the-shelf AR devices usually do not have sufficient computational power (\textit{e.g.,} GPU) to support real-time CV. To enable affordance segmentation, our system streams video data between the AR headset and an external server. However, system latency prevented the overlays from keeping pace with the participants' head motions, negatively impacting their trust in CookAR's intelligence and perceived system usability. To address system latency, we need advancements in both software (\textit{e.g.,} real-time AI models) and hardware (\textit{e.g.,} AR devices with powerful GPUs), which will also increase the overall usability of AI-powered AR systems in dynamic real-world activities.


\subsection{Limitations \& Future Directions}
We outline four primary limitations in this work. First, as an initial prototype, we conducted a qualitative user study with a relatively small number of participants to explore usability and solicit reactions to AR-based affordance augmentations. Future work should conduct larger scale studies with more participants and diverse visual conditions. Second, we re-emphasize the aforementioned technical challenges and limitations. Third, the current CookAR system provides only one basic affordance augmentation---solid-colored overlays. Building upon the design insights in our study, future work should incorporate more augmentation options (\textit{e.g.,} outlines) and enable more flexible adjustments (\textit{e.g.,} colors, thickness of the outlines) to provide LV users more personalized experience.
\rev{Finally, current CookAR system focuses on leveraging CV methods to detect affordances. While some suggested affordances, such as entrance, exit, and activation areas, can be achieved by dataset extension and model fine-tuning, others may not. For example, detecting heated areas may require a thermal sensor and identifying intersection areas can benefit from a LiDAR sensor. Future research should consider additional sensors beyond RGB cameras.}



\section{Conclusion}
In this paper, we introduce CookAR, a wearable AR system that overlays affordance augmentations in real-time to support safe and efficient kitchen tool interactions for people with low vision. To build CookAR, we assembled an egocentric kitchen tool affordance dataset, fine-tuned an RTMDet-Ins-l model on our dataset (\textit{i.e.,} RTMDet-Ins-l-Cook), and created an AR system with a stereo depth camera to generate real-time affordance augmentations in 3D space. We evaluated CookAR in a three-part lab study with 10 LV participants. Findings indicate participants' preferences for affordance augmentations over whole object augmentations for tool interactions, as well as revealing seven types of tool affordances and corresponding augmentation designs preferred by LV users. Our work highlights the promise of affordance augmentations in supporting hand-object interactions for LV people and advances state-of-the-art AI-powered AR technology as low vision aids.

\begin{acks}
This work was supported by an NSF Graduate Research Fellowship and NSF CHS \#1763199. We also thank Yang Li, Sieun Kim, and XunMei Liu for their help with our open-sourced content.
\end{acks}

\bibliographystyle{ACM-Reference-Format}
\bibliography{main}

\appendix 
\label{sec:appendix}

\section{Part 1 Protocol}
\rev{In Part 1, participants compared real-world baseline (\textit{i.e.,} using their typical method in daily life), augmentation baseline (\textit{i.e.,} whole object augmentations), and CookAR (\textit{i.e.,} affordance augmentations) in a tool-grabbing task. We asked participants Likert questions and open-ended questions after each condition. After Part 1, we asked additional qualitative questions.}

\subsection{\rev{7-Point Likert Scale Questionnaire}}
\begin{enumerate}
    \item[1.] \rev{How effective is the system you just used? Why?}
    \item[2.] \rev{How comfortable are you with seeing these visualizations? Why?}
    \item[3.] \rev{How distracting are these visualizations? Why?}
\end{enumerate}

\subsection{\rev{Post-Condition Open-Ended Questions}}
\begin{enumerate}
    \item[1.] \rev{Please describe your overall experience with the visual augmentations.}
    \item[2.] \rev{What did you like about this system?}
    \item[3.] \rev{What did you dislike about this system?}
    \item[4.] \rev{(If any) Ask participants about any interactions we observed to be unconventional (\textit{e.g.,} grabbing a knife by the blade)}
    \item[5.] \rev{How can this system be improved?}
    \item[6.] \rev{In what other scenarios do you think this system would be useful?}
    \item[7.] \rev{Do you have any additional comments about this system that we failed to capture?}
\end{enumerate}

\subsection{\rev{Post-Part Open-Ended Questions}}
\begin{enumerate}
    \item[1.] \rev{Can you compare your experience completing this study task with and without our research prototypes?}
    \item[2.] \rev{Which condition do you prefer to use? Why?}
    \item[3.] \rev{How can this system be improved? Are there any design recommendations you’d like to make?}
    \item[4.] \rev{In which scenarios do you think this system would be useful?}
    \item[5.] \rev{Do you have any additional comments about this study or the systems?}
\end{enumerate}

\section{Part 2 Protocol}
\rev{In Part 2, participants completed a full cooking task using CookAR. They then answered open-ended questions about their experience.}

\begin{enumerate}
    \item[1.] \rev{Please describe your overall experience cooking with this system.}
    \item[2.] \rev{What did you like about this system?}
    \item[3.] \rev{What did you dislike about this system?}
    \item[4.] \rev{What are some improvements you want to make to the system?}
    \item[5.] \rev{Are there any design recommendations you would like to make?}
    \item[6.] \rev{In which scenarios do you think this system would be useful?}
    \item[7.] \rev{Any additional comments about this study or this system?}
\end{enumerate}

\section{Part 3 Protocol}
\rev{In Part 3, participants brainstormed future designs and applications of CookAR. We asked open-ended questions to guide them.}

\begin{enumerate}
    \item[1.] \rev{Can you think of your own design that would improve the usability of our prototype?}
    \item[2.] \rev{In what other scenarios besides cooking do you think this kind of a system can be applicable?}
    \item[3.] \rev{Any additional comments about the system and the entire user study?}
\end{enumerate}

\end{document}